\documentclass[showpacs,english,aps,prc,twocolumn,floatfix]{revtex4}
\usepackage[T1]{fontenc}
\usepackage[latin1]{inputenc}
\usepackage{graphicx}
\usepackage{amssymb}
\usepackage{verbatim}
\usepackage{epic,eepic}
\usepackage{babel}

\newcommand{\ba}{\begin{eqnarray}}
\newcommand{\ea}{\end{eqnarray}}

\begin{document}
\pagestyle{plain}

\title{Flavor asymmetry of the nucleon 
\footnote{Invited talk at XXXI Symposium on Nuclear Physics, 
Hacienda Cocoyoc, Morelos, Mexico, January 7-10, 2008}}
\author{R. Bijker}
\email{bijker@nucleares.unam.mx}
\affiliation{Instituto de Ciencias Nucleares, 
Universidad Nacional Aut\'onoma de M\'exico, \\
A.P. 70-543, 04510 M\'exico, D.F., M\'exico}

\author{E. Santopinto}
\email{santopinto@ge.infn.it}
\affiliation{I.N.F.N. and Dipartimento di Fisica, via Dodecaneso 33,
Genova, I-16146, Italy}

\date{Recibido el 31 de marzo de 2008, aceptado el 31 de mayo de 2008}

\begin{abstract}
The flavor asymmetry of the nucleon sea is discussed in an unquenched 
quark model for baryons in which the effects of quark-antiquark pairs 
($u \bar{u}$, $d \bar{d}$ and $s \bar{s}$) are taken into account in an 
explicit form. The inclusion of $q \bar{q}$ pairs leads automatically to 
an excess of $\bar{d}$ over $\bar{u}$ quarks in the proton, in ageement 
with experimental data. 

\


\

Se discute la asimetr{\'{\i}}a de sabor del nucle\'on en una extensi\'on 
del modelo de cuarks en que se toman en cuento de manera expl{\'{\i}}cita 
los efectos de creaci\'on de pares cuark-anticuark ($u \bar{u}$, $d \bar{d}$ 
y $s \bar{s}$). La inclusi\'on de los pares $q \bar{q}$ lleva inmediatamente 
a un exceso de cuarks $\bar{d}$ sobre $\bar{u}$ en el prot\'on, de acuerdo 
con los datos experimentales. 

\

\noindent
Descriptores: Protones y neutrones, modelos de cuarks, simetr{\'{\i}}a de sabor, 
regla de suma de Gottfried
\end{abstract}


\maketitle                   

\section{Introduction}

The flavor content of the nucleon 
sea provides an important test for models of nucleon structure. A flavor 
symmetric sea leads to the Gottfried sum rule $S_G=1/3$ \cite{gsr}, whereas 
any deviation from this value is an indication of the $\bar{d}/\bar{u}$ 
asymmetry of the nucleon sea. The first clear evidence of a violation of 
the Gottfried sum rule came from the New Muon Collaboration (NMC) \cite{nmc}, 
which was later confirmed by Drell-Yan experiments \cite{DY} and a 
measurement of semi-inclusive deep-inelastic scattering \cite{hermes}. 
All experiments show evidence that there are more $\bar{d}$ quarks in the 
proton than there are $\bar{u}$ quarks. The experimental results and 
theoretical ideas on the flavor asymmetric sea are summarized in several 
review articles \cite{review}.

\begin{figure}[b]
\centering
\resizebox{0.45\textwidth}{!}{\includegraphics{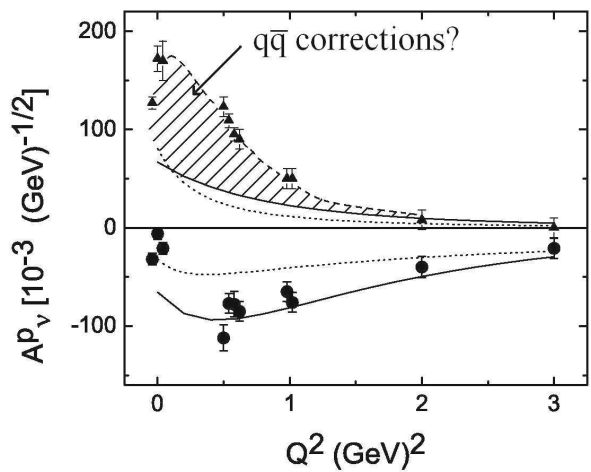}}
\caption[]{Transition form factors for the $N(1520)$ resonance. 
Experimental data are compared with theoretical predictions from the collective 
$U(7)$ model \cite{bil} (dotted line) and the hypercentral model \cite{hypercentral} 
(solid line).}
\label{d13}
\end{figure}

In the constituent quark model (CQM), the proton is described in terms 
of a $uud$ valence-quark configuration. Therefore, a violation of the 
Gottfried sum rule implies the existence of higher Fock components (such 
as $uud-q \bar{q}$ configurations) in the proton wave function. Additional 
indications for the importance of multiquark components are provided by 
parity-violating electron scattering experiments, which have shown evidence 
for a nonvanishing strange quark contribution, albeit small, to the charge 
and magnetization distributions of the proton \cite{Acha}, and by CQM 
studies of baryon spectroscopy \cite{cqm1}. Whereas most models 
(see {\it e.g.} \cite{IK,cqm2,bil,hypercentral}) reproduce the 
mass spectrum of baryon resonances reasonably well, they show 
very similar deviations for other properties, such as for example the 
electromagnetic and strong decay widths of $\Delta(1232)$ and $N(1440)$, 
the spin-orbit splitting of $\Lambda(1405)$ and $\Lambda(1520)$, the 
low $Q^2$ behavior of transition form factors, and the large $\eta$ decay 
widths of $N(1535)$, $\Lambda(1670)$ and $\Sigma(1750)$. 
All of these results point towards the need to include exotic degrees 
of freedom ({\em i.e.} other than $qqq$), such as multiquark $qqq-q \bar{q}$ 
or gluonic $qqq-g$ configurations.  
As an illustration we show in Fig.~\ref{d13} the transverse electromagnetic 
transition form factors of the $N(1520)$ resonance for different CQMs. 
The problem of missing strength at low $Q^2$ can be attributed to the lack of 
explicit quark-antiquark degrees of freedom, which become more important in 
the outer region of the nucleon. 

Theoretically, the role of $q^4 \bar{q}$ configurations in the nucleon 
wave function was studied in an application to the electromagnetic 
form factors \cite{riska}. Mesonic contributions to the spin and flavor 
structure of the nucleon are reviewed in \cite{review}. 

\begin{figure}
\centering
\resizebox{0.3\textwidth}{!}{\includegraphics{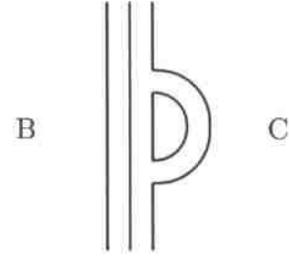}}
\caption{One-loop diagram at the quark level}
\label{fig:1}
\end{figure}

In another, CQM based, approach the importance of $s \bar{s}$ pairs in the 
proton was studied in a flux-tube breaking model based on valence-quark plus 
glue dominance to which $s \bar{s}$ pairs are added in perturbation \cite{baryons}. 
The pair-creation mechanism is inserted at the quark level and the one-loop 
diagrams are calculated by summing over a complete set of intermediate 
baryon-meson states $BC$ (see Fig.~\ref{fig:1}). 
For consistency with the OZI-rule and to retain the success of the CQM in 
hadron spectroscopy, it was found necessary to sum over a complete set of 
intermediate states, including both pseudoscalar and vector mesons, rather 
than just a few low-lying states \cite{baryons,OZI}.  

In order to address the violation of the Gottfried sum rule, we first generalize 
the model of \cite{baryons} to include $u \bar{u}$ and $d \bar{d}$ loops as well. 
The formalism of the ensuing unquenched quark model is reviewed briefly before 
discussing an application to the flavor asymmetry of the nucleon sea. 

\section{Unquenched quark model}

In the flux-tube model for hadrons, the quark potential model arises from an 
adiabatic approximation to the gluonic degrees of freedom embodied in the flux 
tube \cite{flux}. The role of quark-antiquark pairs in meson spectroscopy was 
studied in a flux-tube breaking model \cite{mesons} in which 
the $q \bar{q}$ pair is created with the $^{3}P_0$ quantum numbers of the vacuum.  
Subsequently, it was shown by Geiger and Isgur \cite{OZI} that a {\it miraculous}  
set of cancellations between apparently uncorrelated sets of intermediate states 
occurs in such a way that they compensate each other and do not destroy the good 
CQM results for the mesons. In particular, the OZI hierarchy is preserved and 
there is a near immunity of the long-range confining potential, since the change 
in the linear potential due to the creation of quark-antiquark pairs in the string 
can be reabsorbed into a new strength of the linear potential, {\em i.e.} in a new 
string tension. As a result, the net effect of the mass shifts from pair creation 
is smaller than the naive expectation of the order of the strong decay widths. 
However, it is necessary to sum over  large towers of intermediate 
states to see that the spectrum of the mesons, after unquenching and renormalizing, 
is only weakly perturbed. An important conclusion is that no simple truncation of 
the set of meson loops is able to reproduce such results \cite{OZI}.

\begin{figure}
\centering
\resizebox{0.45\textwidth}{!}{\includegraphics{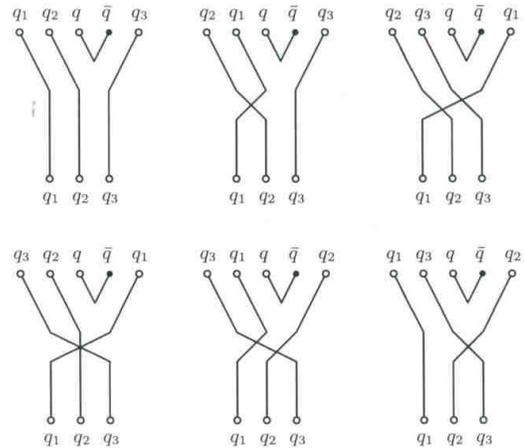}}
\caption{Quark line diagrams for $A \rightarrow B C$ with 
$q_1 q_2 q_3 = uud$ and $q \bar{q} = s \bar{s}$}
\label{diagrams}
\end{figure}

The extension of the flux-tube breaking model to baryons requires a proper treatment 
of the permutation symmetry between identical quarks. As a first step, Geiger and Isgur 
investigated the importance of $s \bar{s}$ loops in the proton by taking into account 
the contribution of the six different diagrams of Fig.~\ref{diagrams} with 
$q \bar{q}=s \bar{s}$ and $q_1 q_2 q_3 = uud$, and by using harmonic oscillator wave 
functions for the baryons and mesons \cite{baryons}. 
In the conclusions, the authors emphasized that, in order to investigate the origin 
of the violation of the Gottfried sum rule and the spin crisis of the proton, it is 
necessary to extend their calculation to include $u \bar{u}$ and $d \bar{d}$ loops as 
well. In this contribution, we take up this challenge and present a generalization 
of the formalism of \cite{baryons} in which quark-antiquark contributions 
can be studied 
\begin{itemize}
\item for any initial baryon resonance, 
\item for any flavor of the quark-antiquark pair, and 
\item for any model of baryons and mesons.
\end{itemize}
These extensions were made possible by two developments: the solution of the 
problem of the permutation symmetry between identical quarks by means of 
group-theoretical techniques, and the construction of an algorithm to 
generate a complete set of intermediate states for any model of baryons 
and mesons. While the first improvement allows the evaluation of the 
contribution of quark-antiquark pairs for any initial baryon $q_1 q_2 q_3$ 
(ground state or resonance) and for any flavor of the $q \bar{q}$ pair 
(not only $s\bar{s}$, but also $u\bar{u}$ and $d\bar{d}$), the second one 
permits the carry out the sum over intermediate states up to saturation for 
any model of baryons and mesons, as long as their wave functions are expressed 
in the basis of harmonic oscillator wave functions. 

The ensuing unquenched quark model is based on an adiabatic treatment of the 
flux-tube dynamics to which $q \bar{q}$ pairs with vacuum quantum numbers are 
added as a perturbation \cite{baryons}. The pair-creation mechanism is 
inserted at the quark level and the one-loop diagrams are calculated by 
summing over a complete set of intermediate states. Under these assumptions, 
to leading order in pair creation, the baryon wave function is given by 
\begin{eqnarray} 
\mid \psi_A \rangle &=& {\cal N} \left[ \mid A \rangle 
+ \sum_{BC l J} \int d \vec{k} \, \mid BC \vec{k} \, l J \rangle \right. 
\nonumber\\
&& \hspace{2cm} \times \left. 
\frac{ \langle BC \vec{k} \, l J \mid T^{\dagger} \mid A \rangle } 
{M_A - E_B - E_C} \right] ~,
\end{eqnarray}
where $A$ denotes the initial baryon and $B$ and $C$ the intermediate 
baryon and meson, $\vec{k}$ and $l$ represent the relative radial momentum and 
orbital angular momentum of $B$ and $C$, and $J$ is the total angular momentum 
$\vec{J} = \vec{J}_B + \vec{J}_C + \vec{l}$. The operator 
$T^{\dagger}$ represents the quark-antiquark pair-creation operator 
with the $^{3}P_0$ quantum numbers of the vacuum \cite{roberts} 
\begin{eqnarray}
T^{\dagger} &=& -3 \sum_{ij} \int d \vec{p}_i \, d \vec{p}_j \, 
\delta(\vec{p}_i + \vec{p}_j) \, C_{ij} \, F_{ij} \, \Gamma(\vec{p}_i - \vec{p}_j) \, 
\nonumber\\
&& \hspace{1cm} \left[ \chi_{ij} \times {\cal Y}_{1}(\vec{p}_i - \vec{p}_j) \right]^{(0)} 
b_i^{\dagger}(\vec{p}_i) \, d_j^{\dagger}(\vec{p}_j) ~.   
\end{eqnarray}
Here, $b_i^{\dagger}(\vec{p}_i)$ and $d_j^{\dagger}(\vec{p}_j)$ are the 
creation operators for a quark and antiquark with momenta $\vec{p}_i$ and 
$\vec{p}_j$, respectively. The quark pair is characterized by a color singlet 
wave function $C_{ij}$, a flavor singlet wave function $F_{ij}$ and a spin 
triplet wave function $\chi_{ij}$ with spin $S=1$. 
The solid harmonic ${\cal Y}_{1}(\vec{p}_i - \vec{p}_j)$ indicates that 
the quark and antiquark are in a relative $P$ wave. 

Since the operator $T^{\dagger}$ creates a pair of constituent quarks, 
a Gaussian quark-antiquark creation vertex function was introduced 
by which the pair is created as a finite object with an effective size, 
rather than as a pointlike object. In momentum space it is given by   
\begin{eqnarray}
\Gamma(\vec{p}_i - \vec{p}_j) = \gamma_0 \, 
\mbox{e}^{-r_q^2 (\vec{p}_i - \vec{p}_j)^2/6} ~. 
\end{eqnarray}
The width has been determined from meson decays to be approximately $0.25-0.35$ 
fm \cite{baryons,OZI,SBG}. Here we take the average value, $r_q=0.30$ fm. 
Finally, the dimensionless constant $\gamma_0$ is the intrinsic pair creation 
strength which has been determined from strong decays of baryons as 
$\gamma_0=2.60$ \cite{CR}. 

The strong coupling vertex
\begin{equation}
\langle B C \vec{k} \, l J \mid T^{\dagger} \mid A \rangle ~,
\end{equation}
was derived in explicit form in the harmonic oscillator basis \cite{roberts}. 
In the present calculations, we use harmonic oscillator wave functions in which 
there is a single oscillator parameter for the baryons and another one for the 
mesons which, following \cite{baryons}, are taken to be $\beta_{\rm baryon}=0.32$ 
GeV \cite{IK} and $\beta_{\rm meson}=0.40$ (GeV) \cite{mesons}, respectively.   

In general, matrix elements of an observable $\hat{\cal O}$ can be expressed as 
\begin{eqnarray}
{\cal O} = \langle \psi_A \mid \hat{\cal O} \mid \psi_A \rangle 
= {\cal O}_{\rm val} + {\cal O}_{\rm sea} ~,
\end{eqnarray}
where the first term denotes the contribution from the valence quarks  
\begin{eqnarray}
{\cal O}_{\rm val} = {\cal N}^2 \langle A \mid \hat{\cal O} \mid A \rangle ~,
\end{eqnarray}
and the second term that from the $q \bar{q}$ pairs
\begin{eqnarray}
{\cal O}_{\rm sea} = {\cal N}^2 \sum_{BC l J,B'C' l' J'} \int d \vec{k} \,
d \vec{k}^{\, \prime} \,
\frac{ \langle A \mid T \mid B' C' \vec{k}^{\, \prime} \, l' J' \rangle } 
{M_A - E_{B'} - E_{C'}} 
\nonumber\\
\langle B' C' \vec{k}^{\, \prime} \, l' J' \mid \hat{\cal O}  
\mid B C \vec{k} \, l J \rangle \, 
\frac{ \langle B C \vec{k} \, l J \mid T^{\dagger} \mid A \rangle } 
{M_A - E_B - E_C} ~.
\label{me}
\end{eqnarray}
We developed an algorithm based upon group-theoretical 
techniques to generate a complete set of intermediate states of good 
permutational symmetry, which makes it possible to perform the sum over 
intermediate states up to saturation, and not just for the first few shells 
as in \cite{baryons}. Not only does this have a significant impact on the 
numerical result, but it is necessary for consistency 
with the OZI-rule and the success of CQMs in hadron spectroscopy. 

\section{Flavor asymmetry}
 
The first clear evidence for the flavor asymmetry of the nucleon sea was provided 
by NMC at CERN \cite{nmc}. The flavor asymmetry is related to the Gottfried integral 
for the difference of the proton and neutron electromagnetic structure functions  
\begin{eqnarray}
S_G &=& \int_0^1 dx \frac{F_2^p(x)-F_2^n(x)}{x} 
\nonumber\\ 
&=& \frac{1}{3} - \frac{2}{3} \int_0^1 dx \left[ \bar{d}(x) - \bar{u}(x) \right] 
\nonumber\\ 
&=& \frac{1}{3} \left[ 1 - 2(N_{\bar{d}}-N_{\bar{u}}) \right] ~.
\end{eqnarray}
Under the assumption of a flavor symmetric sea $\bar{d}(x)=\bar{u}(x)$ one obtains the 
Gottfried sum rule $S_G=1/3$. The final NMC value is $0.2281 \pm 0.0065$ at $Q^2 = 4$ 
(GeV/c)$^2$ for the Gottfried integral over the range $0.004 \leq x \leq 0.8$ \cite{nmc}, 
which implies a flavor asymmetric sea. The violation of the Gottfried sum rule has been 
confirmed by other experimental collaborations \cite{DY,hermes}. Table~\ref{tab:1} 
shows that the experimental values of the Gottfried integral are consistent with each 
other within the quoted uncertainties, even though the experiments were performed 
at very different scales, as reflected in the average $Q^2$ values.  
Theoretically, it was shown that in the framework of the cloudy bag model  
the coupling of the proton to the pion cloud provides a mechanism to produce a flavor 
asymmetry due to the dominance of $n \pi^+$ among the virtual configurations \cite{Thomas}.  

In the unquenched quark model, the flavor asymmetry can be calculated from the 
difference of the number of $\bar{d}$ and $\bar{u}$ sea quarks in the proton 
\begin{eqnarray}
\hat{\cal O} = \hat{N}_{\bar{d}}-\hat{N}_{\bar{u}} ~. 
\label{asym}
\end{eqnarray}
Even in absence of explicit information on the (anti)quark distribution functions, 
the integrated value can be obtained directly from Eq.~(\ref{asym}). The effect of 
the quark-antiquark pairs on the Gottfried integral is a reduction of about one third 
with respect to the Gottfried sum rule, corresponding to an excess of $\bar{d}$ over 
$\bar{u}$ quarks in the proton which is in qualitative agreement with the NMC result. 

An explicit calculation with harmonic oscillator wave functions for the baryons and 
mesons in which the sum over intermediate states includes four oscillator shells, 
shows a proton asymmetry $A_{\rm asym}(p)=N_{\bar{d}}-N_{\bar{u}}=0.21$ which 
corresponds to $S_G=0.19$, in remarkable agreement with the experimental value. 
It is important to note that in this calculation the parameters were taken from 
the literature \cite{baryons,CR}, and that no attempt was made to optimize their 
values. Due to isospin symmetry, the neutron has a similar excess of $\bar{u}$ 
over $\bar{d}$ quarks $A_{\rm asym}(n)=N_{\bar{u}}-N_{\bar{d}}=0.21$. 

\begin{table}[t]
\caption{Experimental values of the Gottfried integral}
\label{tab:1}
\begin{tabular}{lccc}
\hline\noalign{\smallskip}
Experiment & $\langle Q^2 \rangle$ & $x$ range & $S_G$ \\
\noalign{\smallskip}\hline\noalign{\smallskip}
NMC & $4$ & $0.004 < x < 0.80$ & $0.2281 \pm 0.0065$ \\
HERMES & $2.3$ & $0.020 < x < 0.30$ & $0.23 \pm 0.02$ \\
E866/NuSea & $54$ & $0.015 < x < 0.35$ & $0.255 \pm 0.008$ \\
\noalign{\smallskip}\hline
\end{tabular}
\end{table}

\section{Summary and conclusions}

In this contribution, we discussed the importance of quark-antiquark pairs 
in baryon spectroscopy. To this end, we developed an unquenched quark model 
for baryons in which the contributions from $u \bar{u}$, $d \bar{d}$ and 
$s \bar{s}$ loops are taken into account in a systematic way. The present 
model is an extension of the flux-tube breaking model of Geiger and Isgur 
\cite{baryons}, and is valid for any initial baryon resonance, any flavor 
of the quark-antiquark pair and any model of baryons and mesons.

The model was applied to the flavor asymmetry of the nucleon sea. In a first 
calculation with harmonic oscillator wave functions for both baryons and mesons 
in which the parameters were taken from the literature \cite{baryons,CR}, 
it was shown that the inclusion of $q \bar{q}$ pairs leads automatically 
to an excess of $\bar{d}$ over $\bar{u}$ quarks in the proton. The value 
that we obtained for the violation of the Gottfried sum rule is in amazing 
agreement with the experimental data. We emphasize that no attempt was made 
to optimize the parameters in the calculations. 

The first applications of the unquenched quark model to the flavor asymmetry 
of the nucleon sea and the proton spin \cite{BS1} are very promising and encouraging. 
We believe that the inclusion of the effects of quark-antiquark pairs in a 
general and consistent way may provide a major improvement to the constituent 
quark model, which increases considerably its range and applicability. 
In future work, the present unquenched quark model will be applied systematically 
to several problems in light baryon spectroscopy, such as the electromagnetic and 
strong couplings, the elastic and transition form factors of baryon resonances, 
their sea quark content and their flavor decomposition \cite{BS2}.  

\section*{Acknowledgments}
This work was supported in part by PAPIIT-UNAM, Mexico (grant IN113808)  
and in part by I.N.F.N., Italy.

\end{document}